\documentclass[aps]{revtex4}
\usepackage{graphicx}
\usepackage{dcolumn}
\usepackage{bm}
\usepackage{epstopdf}
\usepackage{mathtools}
\usepackage{natbib}
\usepackage{captcont}

\begin{document}

\title{Error Assessment in Modeling with Fractal Brownian Motions}
\author{
Bingqiang Qiao and Siming Liu
}
\affiliation{Key Laboratory of Dark Matter and Space Astronomy, Purple Mountain Observatory, Chinese Academy of Sciences, Nanjing, 210008, China, liusm@pmo.ac.cn}

\begin{abstract}
{
To model a given time series $F(t)$ with fractal Brownian motions (fBms), it is necessary to have appropriate error assessment for related quantities. Usually the fractal dimension $D$ is derived from the Hurst exponent $H$ via the relation $D=2-H$, and the Hurst exponent can be evaluated by analyzing the dependence of the rescaled range $\langle|F(t+\tau)-F(t)|\rangle$ on the time span $\tau$. For fBms, the error of the rescaled range not only depends on data sampling but also varies with $H$ due to the presence of long term memory. This error for a given time series then can not be assessed without knowing the fractal dimension. We carry out extensive numerical simulations to explore the error of rescaled range of fBms and find that for $0<H<0.5$, $|F(t+\tau)-F(t)|$ can be treated as independent for time spans without overlap; for $0.5<H<1$, the long term memory makes $|F(t+\tau)-F(t)|$ correlated and an approximate method is given to evaluate the error of $\langle|F(t+\tau)-F(t)|\rangle$. The error and fractal dimension can then be determined self-consistently in the modeling of a time series with fBms.
}
\end{abstract}

\maketitle

\section{Introduction}

Fractal Brownian motions (fBms) are a very useful tool to better understand many natural time series \cite{m68, p78, l99}. The fractal dimension $D$ is one of the most important quantities to be evaluated for any given time series $F(t)$ and is related to the Hurst exponent $H$ via $D=2-H$. The Hurst exponent $H$ can be defined with the following expression:
\begin{equation}
\langle|F(t+\tau)-F(t)|\rangle \propto \tau^H\,,
\end{equation}
where $^\prime\langle\rangle^\prime$ represents averaging over time $t$. In practice, for a given duration of a time series $T$, appropriate error needs to be assigned to the rescaled range $\langle|F(t+\tau)-F(t)|\rangle$ to obtain a quantitative measurement of $H$. But the correlation of fBms makes the error depend not only on the sampling but also on the $H$, which is not known as a prior. The error and $H$ need to be determined simultaneously in a self-consistent modeling.

In principle, one may simulate the sampling of $F(t)$ for assumed $H$ of an fBm and study the statistical similarity of the simulated result and $F(t)$. The best fit value of $H$ is obtained when this statistical similarity reaches the maximum. However, this process can be cumbersome in practical applications \cite{c09}. It is therefore necessary to explore the characteristics of the rescaled range of fBms so that some approximate method may be derived to assess the related error efficiently. In this paper, we first study the standard deviation of the mean of increments $\langle F(t+\tau)-F(t)\rangle$ for a given $\tau$ for several sampling methods (Section 2). In Section 3, a similar study is carried out on the rescaled range $\langle |F(t+\tau)-F(t)|\rangle$. Possible application of these results and conclusions are given in Section 4.

\section{Standard Deviation of $\langle F(t+\tau)-F(t)\rangle$}

\begin{figure*}[htb]
\centering
\includegraphics[height=54mm,angle=0]{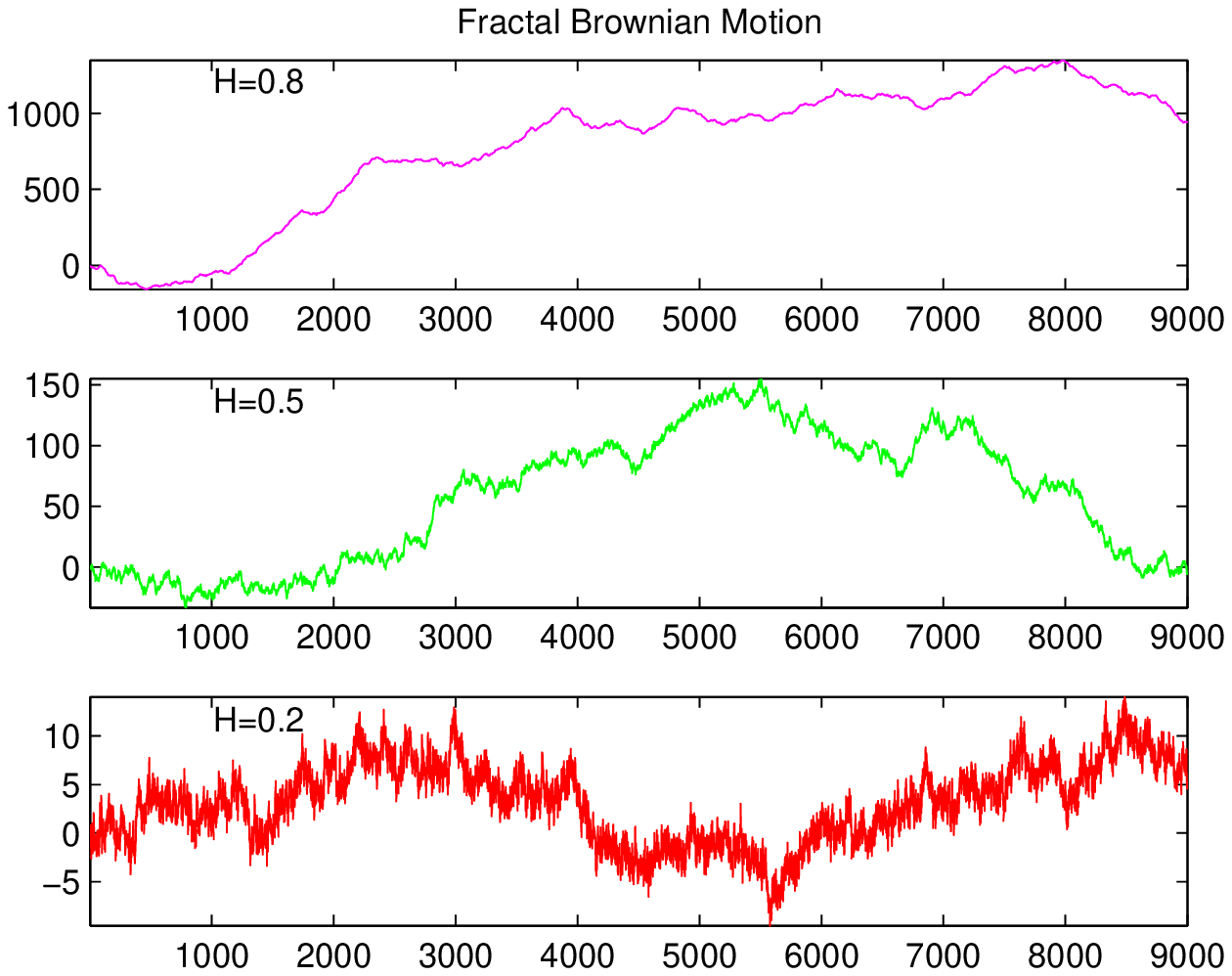}
\includegraphics[height=55mm,angle=0]{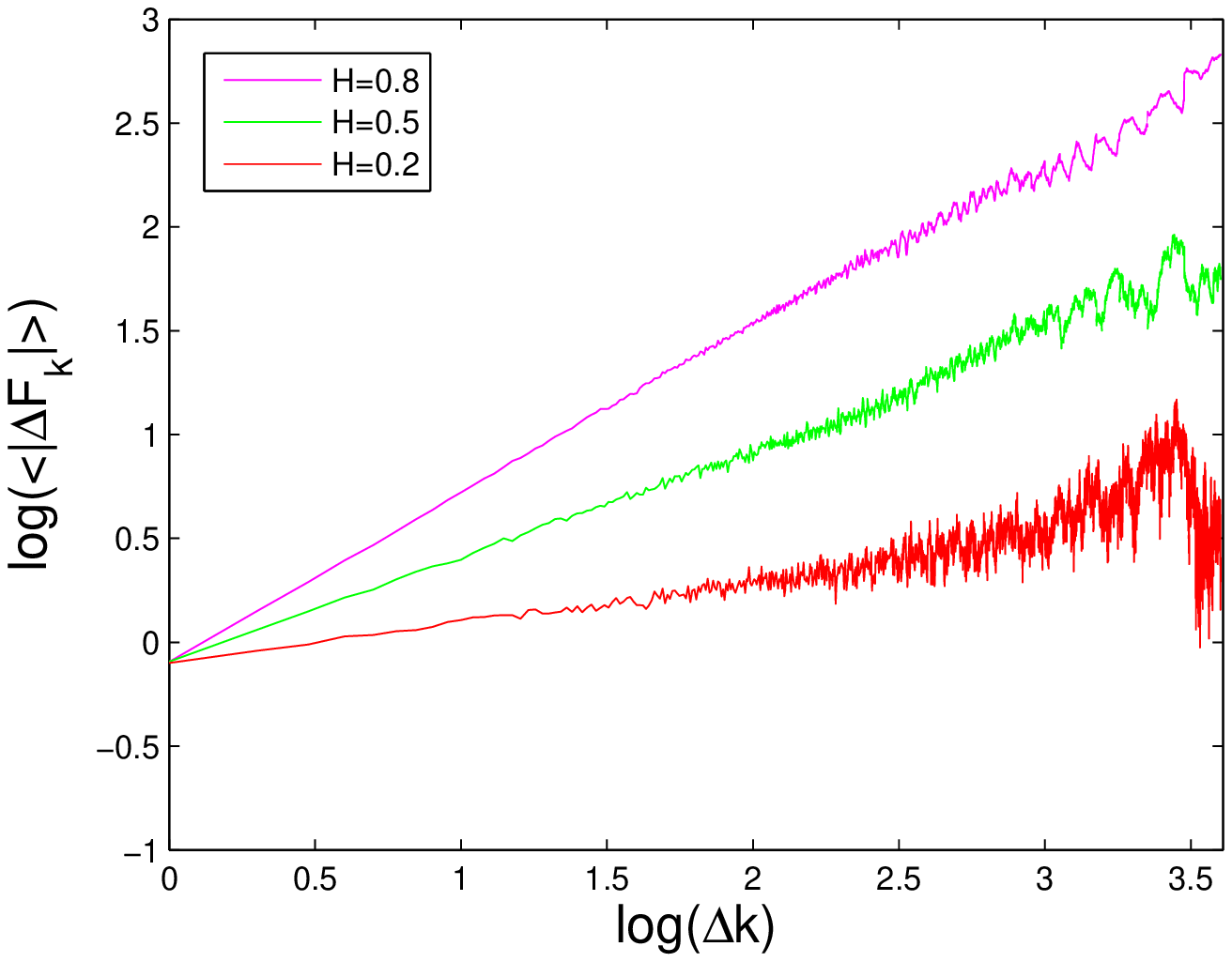}
\caption{Left: A few examples of fBms. The Hurst exponent is indicated on the figure. Right: Dependence of the rescaled range $\langle|\Delta F_k|\rangle$ on the time span $\Delta k$ for the examples in the left panels.}
\label{f1}
\end{figure*}

In the following, we will use discrete-time fBms $F(k)$ with $k$ representing the time steps, $\Delta k$ the time span, and $F(0)=0$. We use the Lowen method to generate fBms with $0<H<0.5$ and the circulant embedding method for $0.5<H<1$. Figure \ref{f1} shows a few examples of fBms with different $H$. The right panel shows the dependence of the rescaled range $\langle |\Delta F_k|\rangle =\langle |F(k+\Delta k)-F(k)|\rangle$ on the time span $\Delta k$. In this case we have a continuous sampling with
\begin{equation}
\langle |\Delta F_k|\rangle = {\displaystyle\sum_{i=0}^{{\rm Int}(N/\Delta k)-1} |F([i+1]*\Delta k)-F(i*\Delta k)|\over {\rm Int} (N/\Delta k)} \,,
\end{equation}
where Int$(N/\Delta k)$ rounds $(N/\Delta k)$ to its integer part.
 We see the slope of the $\log\langle|\Delta F_k|\rangle$---$\log(\Delta k)$ plot is consistent with $H$
 and the fluctuation of $\langle |\Delta F_k|\rangle$ increases with the decrease of the sampling number Int$(N/\Delta k)$. Theoretically, we expect
 $$\langle|\Delta F_k|\rangle =\left({2\over \pi}\right)^{1/2} (\Delta k)^H$$
 An error needs to be assigned to each $\langle |\Delta F_k|\rangle$ to check the consistency of the fBm generation code quantitatively.

The error of $\langle |\Delta F_k|\rangle$ equals to its standard deviation. For a given $H$, one can generate many discrete-time fBms and obtain the distribution and standard deviation of $\langle |\Delta F_k|\rangle$ directly. In this section, we first study the standard deviation of
$\langle \Delta F_k\rangle$. Given the well-defined autocorrelation of fBms \cite{m68, l99},
\begin{equation}
E(F(k)F(m))=(k^{2H}+m^{2H}-|k-m|^{2H})/2 \,,
\label{e3}
\end{equation}
where $E(F(k)F(m))$ represents the expectation of $F(k)F(m)$,
analytical expression for the standard deviation of $\langle \Delta F_k\rangle$ can also be derived to check for consistency with the numerical results.

\begin{figure*}[htb]
\centering
\includegraphics[height=54mm,angle=0]{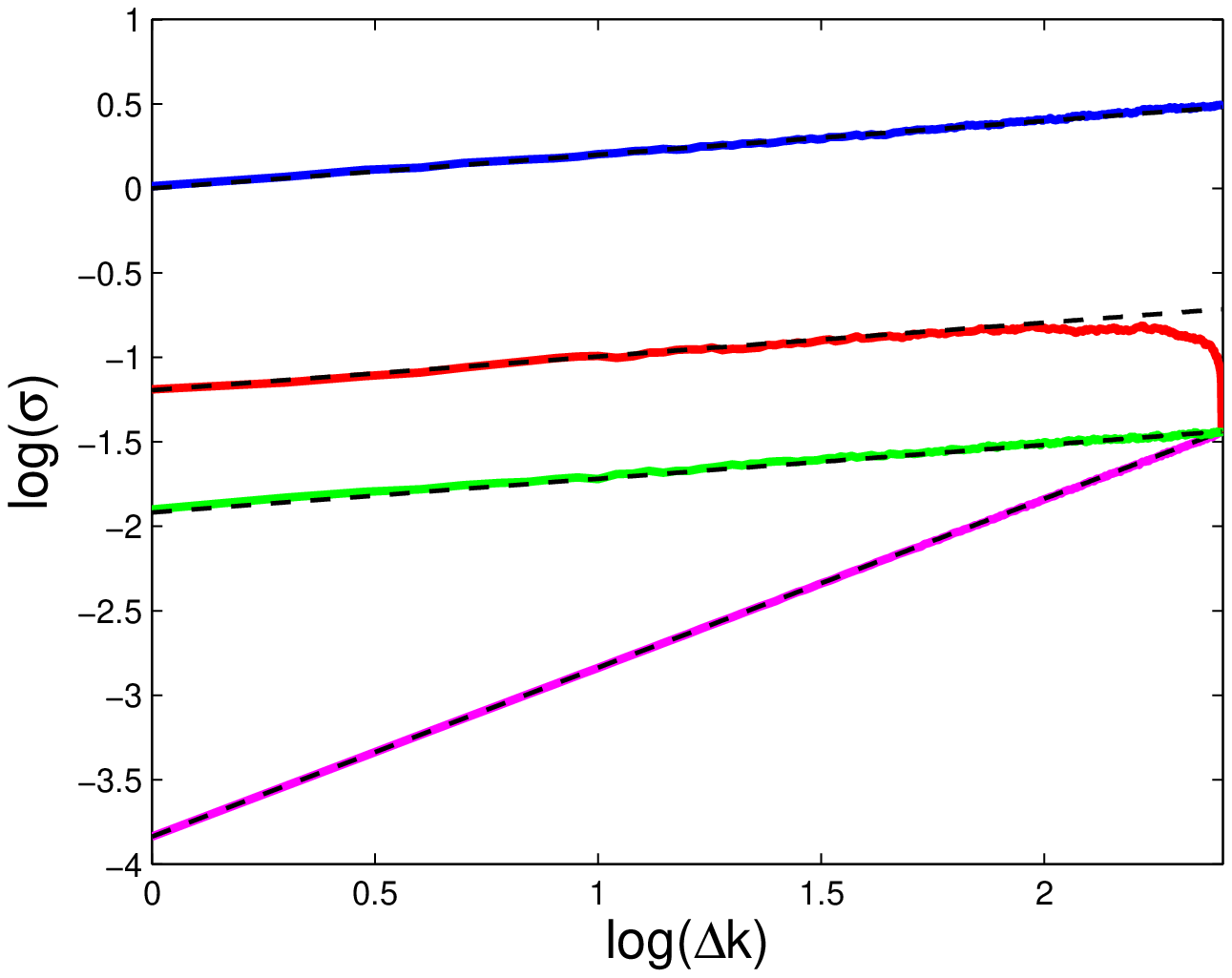}
\includegraphics[height=54mm,angle=0]{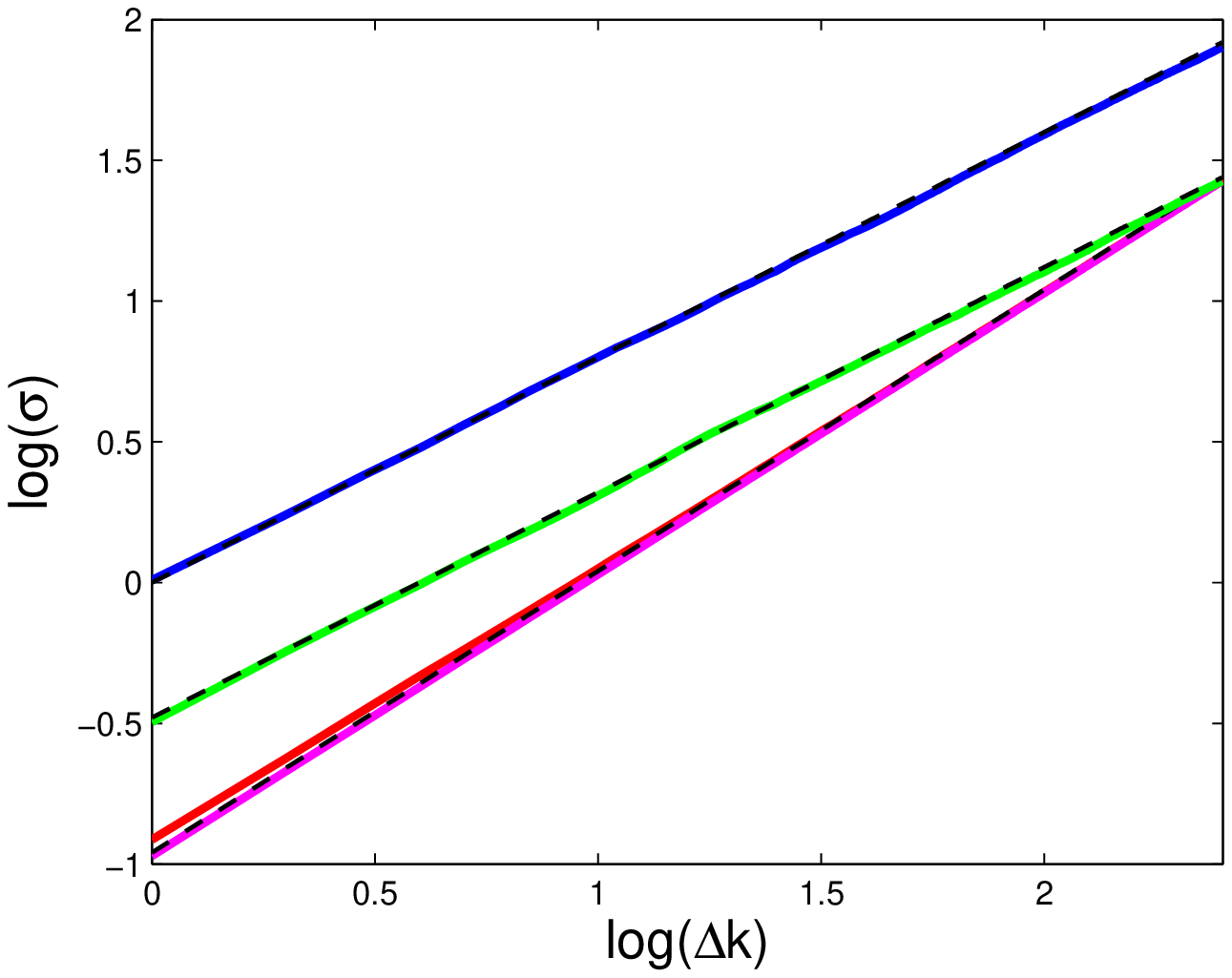}
\caption{
Left: Dependence of the standard deviation of $\langle\Delta F_k\rangle$ on $\Delta k$ for 1000 segments of fBm with $H = 0.2$ and a length of $2^{16}$.  Different colors represent results obtained with different sampling method. Dashed lines indicate the theoretically expected lines. See text for details.
Right: Same as the left panel but for $H=0.8$.
}
\label{f2}
\end{figure*}

Figure 2 shows the dependence of the standard deviation of $\langle \Delta F_k\rangle$ on $\Delta k$. Here we generate 1000 segments of fBm with a length of $2^{16}$ for each $H$. The blue line (case 1) corresponds to the case where only one value of $\Delta F_k=F(\Delta k) - F(0)$ is taken for each segment and
\begin{equation}
\langle\Delta F_k\rangle=F(\Delta k) - F(0)\,.
\end{equation}
As expected, the standard deviation of $\langle\Delta F_k\rangle$ $\sigma$ scales as
$(\Delta k)^{H}$:
$$\sigma = (E(\langle\Delta F_k\rangle^2))^{1/2} =(E( [F(\Delta k)]^2))^{1/2}=(\Delta k)^H\,,$$
where we have used the fact that $E(\langle\Delta F_k\rangle)=0$.

For the red line (case 2), we have
\begin{equation}
\langle\Delta F_k\rangle={\displaystyle\sum_{i=0}^{N-1} [F(\Delta k+N*i) - F(N*i)]\over N}
\end{equation}
with $N=250$. Here we notice that
\begin{eqnarray}
&&E([F(\Delta k+N*i) - F(N*i)][F(\Delta k+N*j) - F(N*j)]\nonumber \\
&=& {\left|1+{\Delta k\over N(i-j)}\right|^{2H}+\left|1-{\Delta k\over N(i-j)}\right|^{2H}-2\over 2|N(i-j)|^{-2H}}\nonumber \\
&\simeq& H(2H-1)(\Delta k)^2|N(i-j)|^{2H-2} \ \ \ \  {\rm for}\ \ \ \  \Delta k\ll |N(i-j)|\,.
\end{eqnarray}
For $H\le0.5$, $E([F(\Delta k+N*i) - F(N*i)][F(\Delta k+N*j) - F(N*j)]\ll (\Delta k)^2|N(i-j)|^{2H-2}$ for $i\ne j$. When $\Delta k\ll N$, we have $E([F(\Delta k+N*i) - F(N*i)][F(\Delta k+N*j) - F(N*j)]\ll (\Delta k)^{2H}=\sigma^2$. One therefore can ignore the correlation between $[F(\Delta k+N*i) - F(N*i)]$ and the standard deviation is reduced by a factor of $N^{1/2}$ as indicated by the dashed line. When $\Delta k$ becomes comparable to $N$, the anti-correlation between $[F(\Delta k+N*i) - F(N*i)]$ is important and the standard deviation decreases with the increase of $\Delta k$.
For $H\ge 0.5$, the correlation between $[F(\Delta k+N*i) - F(N*i)]$ is important and there is no simple analytical expression for $\sigma$ and we just show the numerical results in the right panel.

For the green line (case 3), we have
\begin{equation}
\langle\Delta F_k\rangle={\displaystyle\sum_{i=0}^{N-1} [F(\Delta k+\Delta k*i) - F(\Delta k*i)]\over N}\,,
\end{equation}
with $N=250$. In this case, we have
\begin{eqnarray}
&&E([F(\Delta k+\Delta k*i) - F(\Delta k*i)][F(\Delta k+\Delta k*j) - F(\Delta k*j)]\nonumber \\
&=&{\left|i-j+1\right|^{2H}+\left|i-j-1\right|^{2H}-2|i-j|^{2H}\over 2}(\Delta k)^{2H}
\label{e8}
\,.
\end{eqnarray}
Then the standard deviation of $\langle\Delta F_k\rangle$ $\sigma$ is given by
\begin{equation}
\sigma = N^{H-1} (\Delta k)^H \,.
\end{equation}

For the magenta line (case 4), we have
\begin{equation}
\langle\Delta F_k\rangle={\displaystyle\sum_{i=0}^{{\rm Int}(N^2/\Delta k)-1} [F(\Delta k+k*i) - F(\Delta k*i)]\over {\rm Int}(N^2/\Delta k)-1}
\end{equation}
with $N=250$. Then using Equation (\ref{e8}), we have
\begin{equation}
\sigma = N^{2(H-1)}\Delta k\,.
\end{equation}

\begin{figure*}[htb]
\centering
\includegraphics[height=54mm,angle=0]{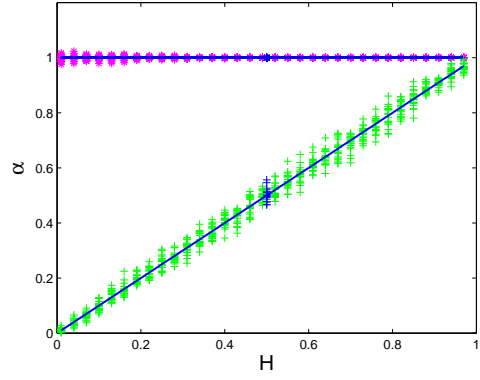}
\caption{
Dependence of the exponent $\alpha$ of the green and magenta lines in Figure \ref{f2} on $H$. Here only 100 segments of fBm are obtained for each $H$. The solid lines indicate the expected lines.}
\label{f3}
\end{figure*}
For cases 2 and 4, $\sigma$ scales with $(\Delta k)^\alpha$ with $\alpha= H$ and $1$, respectively. Figure \ref{f3} shows the dependence of $\alpha$ on $H$. For the numerical results, we did a linear fit to the log$(\sigma)$---log$(\Delta k)$ plot to obtain the exponent $\alpha$.
The above results show that the numerical results are consistent with the theoretical expectations.

\section{Standard Deviation of $\langle |F(t+\tau)-F(t)|\rangle$}

\begin{figure*}[htb]
\centering
\includegraphics[height=54mm,angle=0]{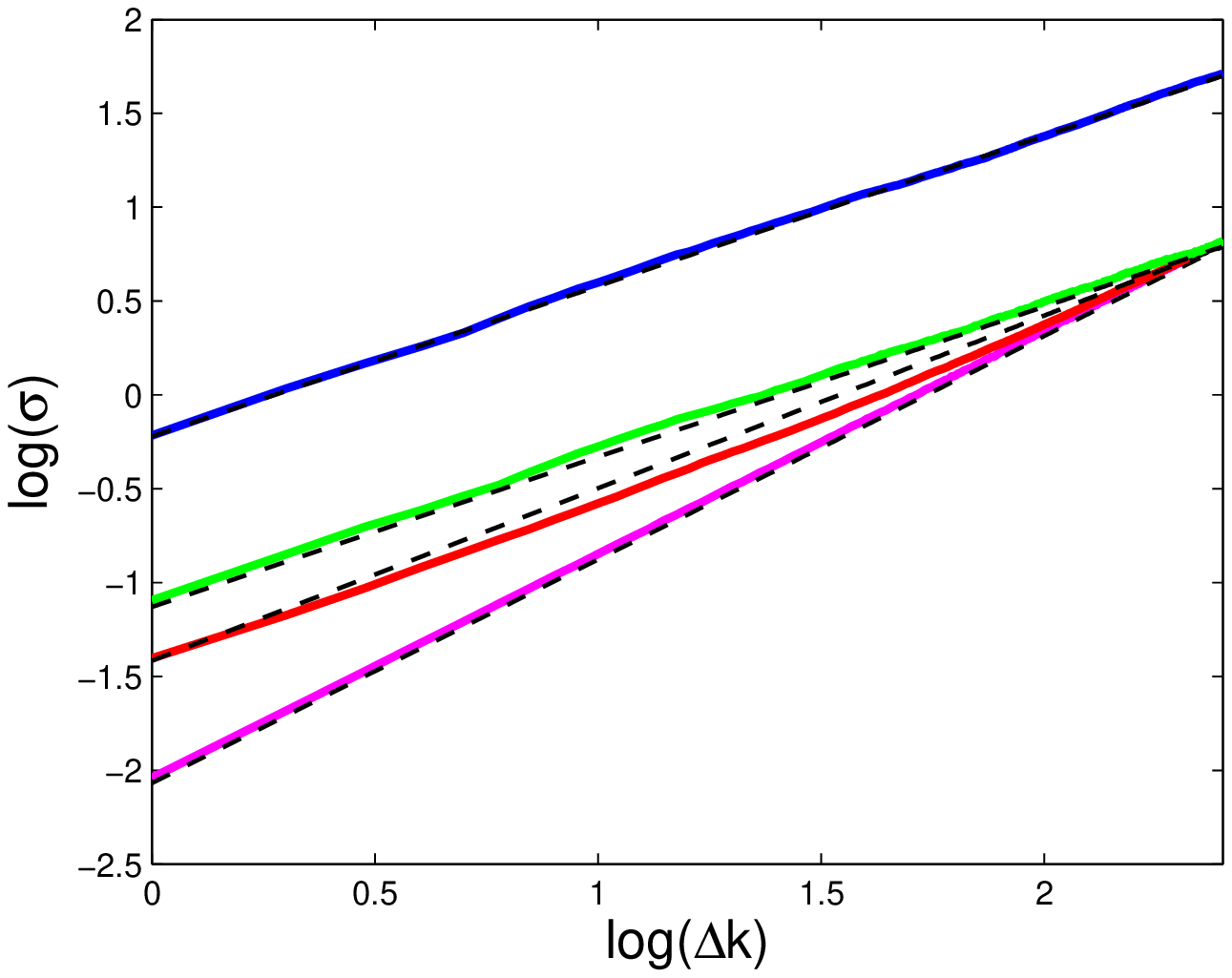}
\includegraphics[height=54mm,angle=0]{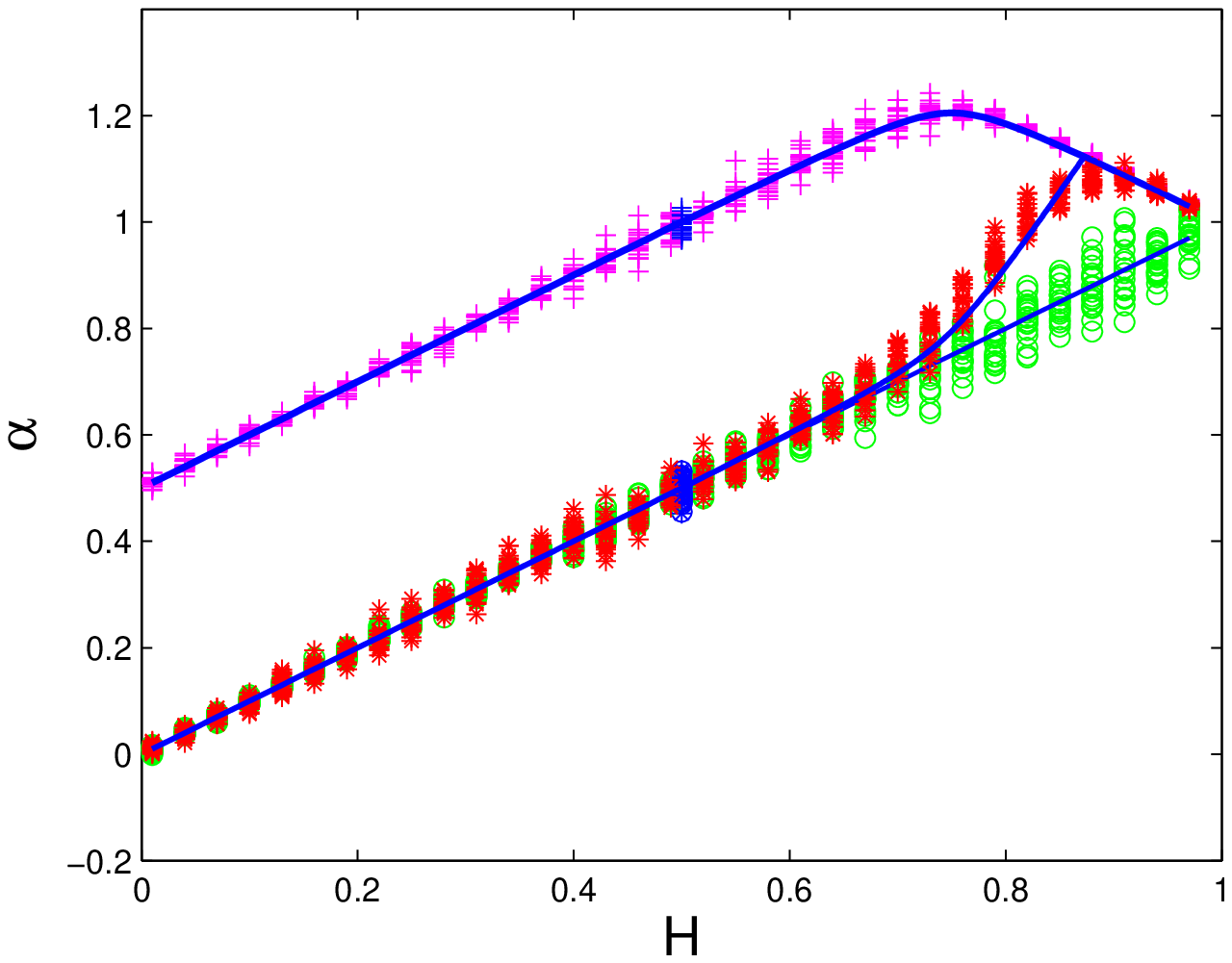}
\caption{
Left: Same as the right panel of Figure \ref{f2} with $H=0.8$ but for the rescaled range $\langle|\Delta F_k|\rangle$ instead of $\langle\Delta F_k\rangle$.
Right: Same as Figure \ref{f3} but for the rescaled range $\langle|\Delta F_k|\rangle$. See text for details.
}
\label{f4}
\end{figure*}

To obtain the Hurst exponent for a given time series, we need to evaluate the standard deviation of $\langle|\Delta F_k|\rangle = \langle |F(k+\Delta k)-F(k)|\rangle$. We carry out a similar analysis as the previous section and the results are shown in Figure \ref{f4}. For the first case (blue line), we expect
\begin{equation}
\sigma = (1-2/\pi)^{1/2}(\Delta k)^H\,.
\end{equation}
For the other cases, since we do not have a simple autocorrelation function like Equation (\ref{e3}) for $\langle\Delta F_k\rangle$, there is no simple analytical expression for the standard deviation $\sigma$. We therefore resort to the empirical approach to derive some approximate expressions based on the numerical results.

First we find that $\langle |F(k+\Delta k)-F(k)|\rangle$ can be treated as independent for $0<H<0.5$ and no overlapping time spans. Then for cases 2 and 3, we have
\begin{equation}
\sigma = N^{-1/2}(1-2/\pi)^{1/2}(\Delta k)^H\,.
\end{equation}
For case 4, we have
\begin{equation}
\sigma = N^{-1}(1-2/\pi)^{1/2}(\Delta k)^{H+0.5}\,.
\end{equation}
Note that the duration of our time series must be greater than $N^2$.

For $0.5<H<1$, we find that $\langle |F(k+\Delta k)-F(k)|\rangle$ for no overlapping spans are not independent. For a given duration of $N^2$ and time span $\Delta k<N^2$, the number of independent rescaled range $In$ can be approximated as
\begin{equation}
 In = {\rm Int} \left[\left({N^2\over \Delta k}\right)^{\beta}\right]\,,
 \end{equation}
 where
 \begin{equation}
 \beta = 2-2H+0.26^{1/2}-[(2H-1.5)^2+0.01]^{1/2}\,.
 \end{equation}
Note that $\beta+2H$ is a hyperbolic function of $H$, $\beta = 1$ for $H=0.5$, $\displaystyle\lim_{H\rightarrow 1}\beta = 0$, and $\beta$ decreases monotonically with the increase of $H$ from 0.5 to 1.0.
 Then for case 3, we have
 \begin{equation}
 \sigma = N^{-\beta/2}(1-2/\pi)^{1/2}(\Delta k)^H\,.
 \end{equation}
For case 4, we have
\begin{equation}
\sigma =  N^{-\beta}(1-2/\pi)^{1/2}(\Delta k)^{H+\beta/2}\,.
\label{e18}
\end{equation}

Case 2 is a bit complicated as in the previous section since $\sigma$ does not scale exactly as $(\Delta k)^\alpha$ with $\alpha$ independent of $\Delta k$. However one may still derive an approximate expression for $\alpha$ by considering the $\sigma$ for $\Delta k = 1$ and $N$. For $\Delta k = 1$, $In ={\rm Int} (N^{2\beta})$. When $N>{\rm Int}(N^{2\beta})$, i.e., $\beta<0.5$, the sample size $N$ is greater than the total number of independent rescaled range for any $\Delta k$. Then we have
\begin{equation}
\sigma = N^{-\beta}(1-2/\pi)^{1/2}(\Delta k)^{H+\beta/2}\,,
\end{equation}
which is the same as Equation (\ref{e18}). When $N<{\rm Int}(N^{2\beta})$, the sampling for $\Delta k=1$ can be considered as independent but the sampling for $\Delta k = N$ is not since ${\rm Int}(N^\beta)\le N$. The standard deviation of $\langle |F(k+\Delta k)-F(k)|\rangle$ at $\Delta k = 1$ and $N$ are respectively $(1-2/\pi)^{1/2}N^{-0.5}$ and $(1-2/\pi)^{1/2}N^{H-\beta/2}$. Then we can get an approximate expression for $\alpha$:
\begin{equation}
\alpha\simeq H-\beta/2+1/2\,.
\end{equation}
Figure \ref{f4} shows that our approximate expressions can fit the numerical results pretty well.

\begin{figure*}[htb]
\centering
\includegraphics[height=54mm,angle=0]{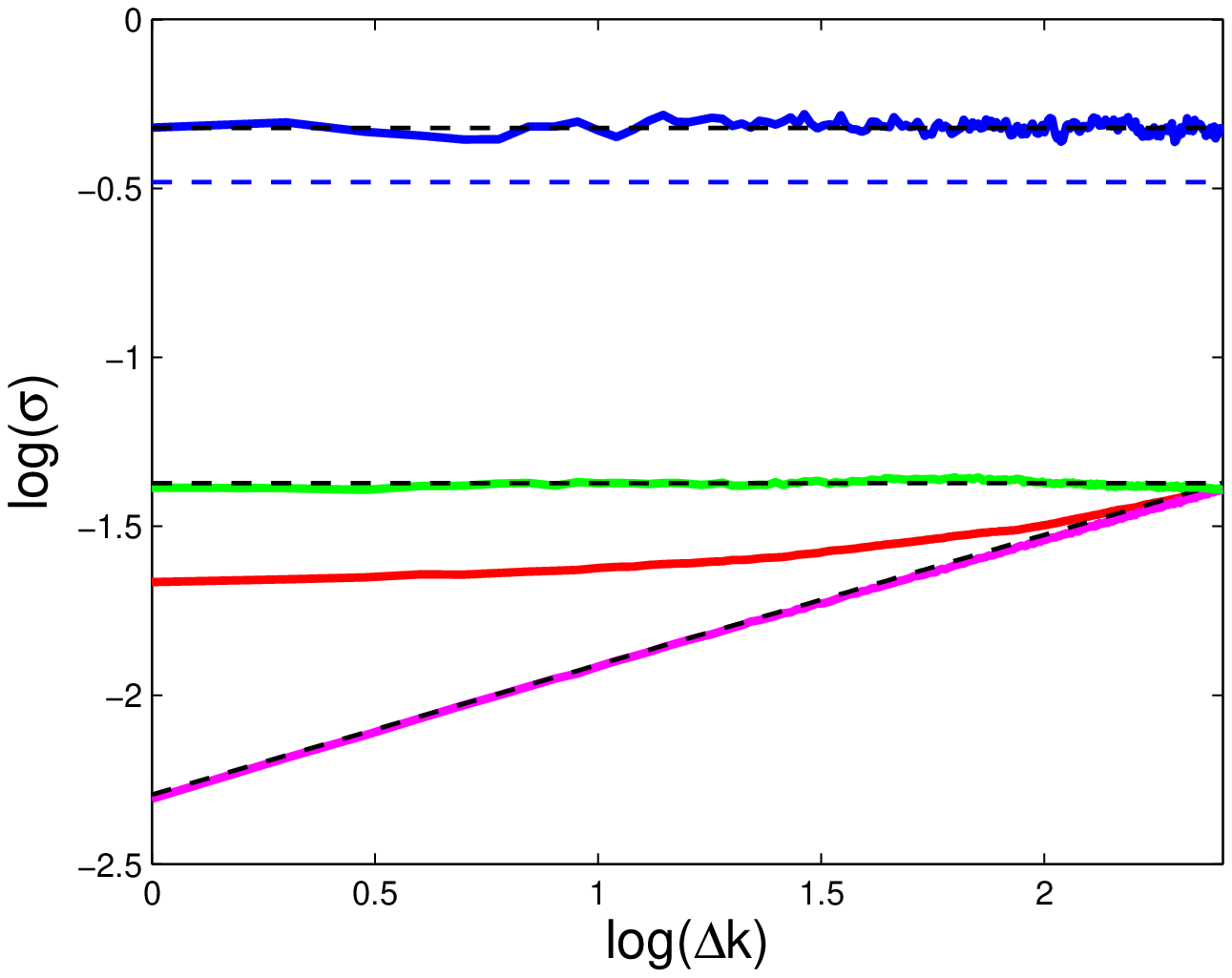}
\includegraphics[height=54mm,angle=0]{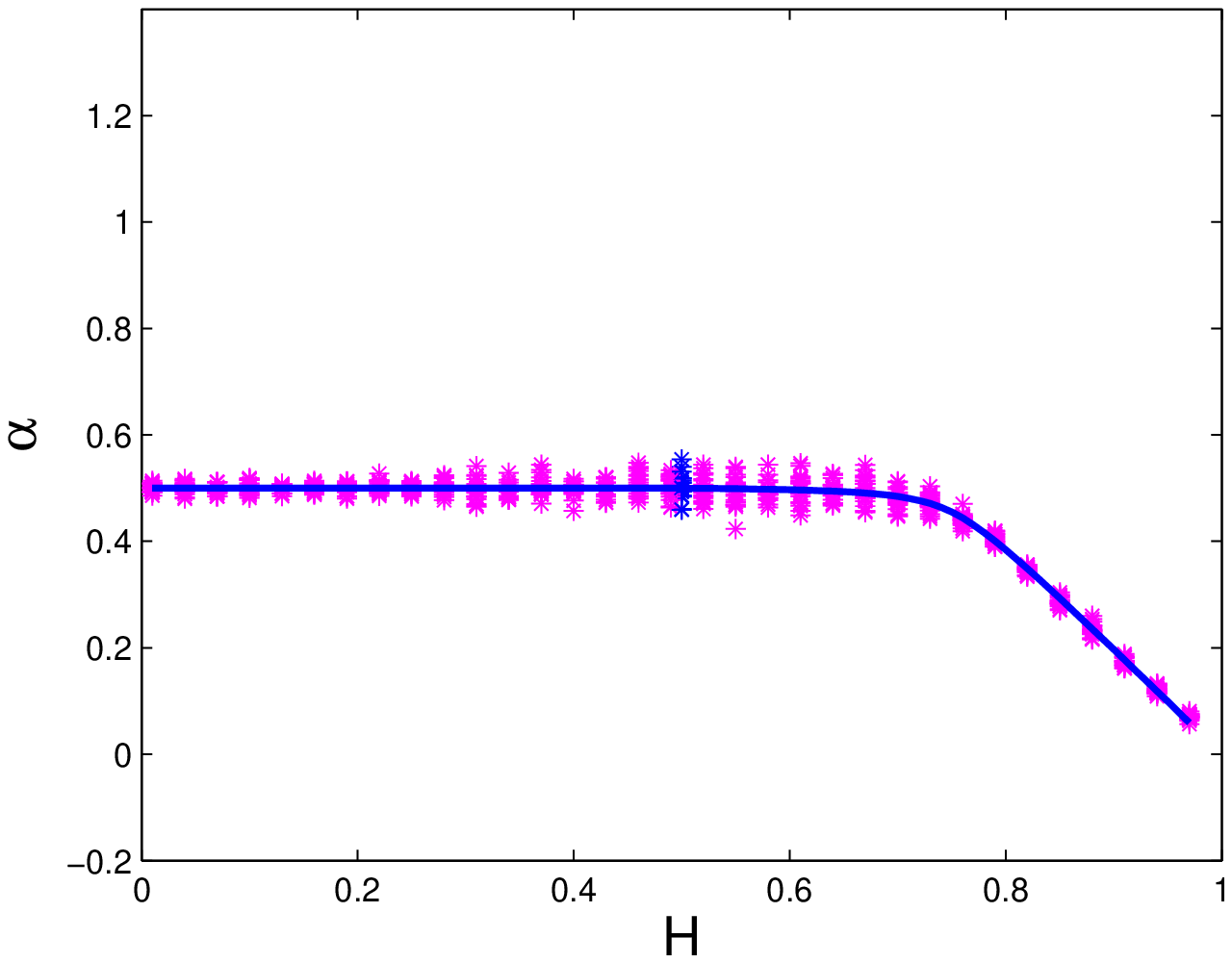}
\caption{
Same as Figure \ref{f4} but for the logarithm of the rescaled range $\langle|\Delta F_k|\rangle$. See text for details.
}
\label{f5}
\end{figure*}

The Hurst exponent is often obtained with a linear fit to the $\log\langle|\Delta F_k|\rangle$---$\log(\Delta k)$ plot. We therefore also study the standard deviation of $\log\langle|\Delta F_k|\rangle$. The results are shown in Figure \ref{f5}. When the standard deviation of $\langle|\Delta F_k|\rangle$ $\sigma_0$ is much smaller than its expectation value, the standard deviation of $\log\langle|\Delta F_k|\rangle$ is given by
\begin{equation}
\sigma = {\log(e)\  \sigma_0 \over \langle|\Delta F_k|\rangle}\,,
\label{e21}
\end{equation}
where $e\simeq 2.71828$ is the base of the natural logarithm.

Then for $0<H<0.5$ and cases 2 and 3, we have
\begin{equation}
\sigma = N^{-1/2}\log(e)\ (\pi/2-1)^{1/2}\,,
\end{equation}
and for case 4, we have
\begin{equation}
\sigma = N^{-1}\log(e)\ (\pi/2-1)^{1/2}(\Delta k)^{0.5}\,.
\end{equation}
For $0.5<H<1$ and case 3, we have
\begin{equation}
 \sigma = N^{-\beta/2}\log(e)\ (\pi/2-1)^{1/2}\,,
 \end{equation}
for case 4, we have
\begin{equation}
\sigma =  N^{-\beta}\log(e)\ (\pi/2-1)^{1/2}(\Delta k)^{\beta/2}\,.
\end{equation}

For case 1, Equation (\ref{e21}) is invalid because the standard deviation of $|\Delta F_k|$ is comparable to its expectation. The blue dotted line in the left panel of Figure \ref{f5} corresponds to $\sigma = \log(e) (\pi/2-1)^{1/2}$ as given by Equation (\ref{e21}). The black dashed line corresponds to the numerically calculated result:
$$
\sigma = \left[\left({2\over \pi}\right)^{1/2} \int_0^\infty (\log(x) -\mu)^2\exp(x^2/2){\rm d} x\right]^{1/2}\,,
$$
where
$$
\mu = \left({2\over \pi}\right)^{1/2} \int_0^\infty \log(x)\exp(x^2/2){\rm d} x\,,
$$
which is consistent with the numerical result.

\section{Conclusion}

In this paper, we investigate the error of the increment and rescaled range of fBms. For the increment, analytical expressions of the error can be derived from the autocorrelation of fBms. There is not simple expression for the error of the rescaled range. However, we find that for $0<H<0.5$, the amplitude of the increment can be treated as independent for no overlapping spans and for $0.5<H<1.0$, the total number of independent amplitude of the increment may be approximated as $(T/\tau)^\beta$ where $T$ is the duration of the fBm and $\tau$ is the time span. We then show that the error of the rescaled range obtained from numerical simulations can be explained with some simple expressions for several sampling methods.

With these results, one can readily assess whether a given time series is consistent with a fBm. To do that, one may obtain the rescaled range as function of time span and estimate the Hurst exponent $H$. With this $H$, error can be assigned to the rescaled range to check the consistency between the time series and an fBm with $H$. The value of $H$ and the error of the rescaled range may also be adjusted to maximum the statistical similarity between the given time series and an fBm.

\section*{Acknowledgements}
This work is partially supported by the NSFC grants 11173064, 11233001, and 11233008.

{}

\end{document}